\documentclass[
 aip,
 rsi,
 amsmath,amssymb,
 reprint,%
]{revtex4-1}

\usepackage{graphicx}
\usepackage{dcolumn}
\usepackage{bm}

\usepackage[utf8]{inputenc}
\usepackage[T1]{fontenc}
\usepackage{mathptmx}
\usepackage{etoolbox}
\usepackage{amsmath}
\DeclareMathOperator{\atantwo}{atan2}
\graphicspath{{Figs/}}
\usepackage{float}
\usepackage[utf8]{inputenc}
\usepackage{color}
\usepackage{nicefrac}

\makeatletter
\def\@email#1#2{%
 \endgroup
 \patchcmd{\titleblock@produce}
  {\frontmatter@RRAPformat}
  {\frontmatter@RRAPformat{\produce@RRAP{*#1\href{mailto:#2}{#2}}}\frontmatter@RRAPformat}
  {}{}
}%
\makeatother
\begin{document}


\title{Ion dynamic characterization using phase-resolved laser-induced fluorescence spectroscopy in a Hall effect thruster}

\author{Y. Dancheva}
\email{The author to whom correspondence may be addressed: yordanka.dancheva@aerospazio.com}
 
\author{P. Coniglio}
\affiliation{Aerospazio Tecnologie S.r.l. - Rapolano Terme, Italy}
 
\author{M. Da Valle}
\affiliation{DSFTA, University of Siena, via Roma 56, Siena, Italy}

\author{F. Scortecci}
\affiliation{Aerospazio Tecnologie S.r.l. - Rapolano Terme, Italy}

\date{\today}

\begin{abstract}
Significant information on the dynamics of the plasma constituents in Hall effect thrusters can be obtained using minimally-intrusive techniques such as laser-induced fluorescence (LIF) diagnostics. Indeed, LIF provides an excellent tool to determine the ion velocity distribution function with high spatial resolution. Even in a steady-state operation, recording time-resolved maps of the velocity distribution is relevant due to persisting  time-dependent features of the thruster discharge. One of the preeminent phenomena that renders the ion velocity distribution time dependent is commonly attributed to the breathing mode, characterized by pronounced oscillations in the discharge current. The goal of this work is to propose a new technique for plasma dynamic studies based on LIF spectroscopy with phase-resolution during the breathing period. 
 To this purpose, the Hilbert transform is used to define the instantaneous phase of oscillation of the thruster current. Ion velocity distribution modification, over assigned phases of oscillation,  is measured simultaneously and in real-time thanks to fully numerical analysis of the data.

\end{abstract}

\maketitle

\section{\label{intro}Introduction}
Hall effect thrusters (HET) are well-studied {\bf E $\times$ B} devices that ionize and accelerate propellant ions and find application in missions such as satellite station keeping, orbit raising and deep space travel, because of their good thrust and high specific impulse, hence low propellant consumption. These systems are overabundant in plasma instabilities and fluctuations with a wide spectral distribution, ranging from 1$\nobreakspace$kHz to 60$\nobreakspace$MHz, \cite{Choueiri_2004, Zhurin_1993} of which many still lack a comprehensive understanding. These phenomena are thought to be critical to drive electron transport across magnetic field lines, and contribute to the propellant ionization and overall thruster performance. To improve understanding and modelling of thruster performance it is often preferable to capture the time-varying characteristics within an oscillation period. Studies on the spatially and time-dependent ion velocity distribution function (IVDF), that can directly impact the performance of plasma systems,  give insight into ionization mechanisms, electric potential formation, and acceleration regions. In addition, the impact of the life-time limiting effects caused by erosion can  be also evaluated. \cite{Bareilles_2004, Chaplin_2019}

The most common and dominant plasma oscillation in HET, often referred as the {\it breathing mode}, is closely related to the propellant ionization (and eventual ion acceleration) and occurs spontaneously in a quasi-periodic process with  frequency typically in the range of 10$\div30\nobreakspace$kHz. Breathing
mode models suggest the presence of a propagating ionization front traversing the channel and giving rise to intense  discharge current oscillations.  J.M.Fife,  S.Barral, and co-workers  proposed an explanation of this phenomenon based on a predator-prey model, which describes it in terms of ion acceleration causing plasma depletion followed by neutrals replenishment. \cite{Fife_1997, Barral_2009} Time-averaged measurement of these processes is not appropriate to reveal the complex physics underlying the operation of these devices. However, resolving fluctuating properties at time scales of tens of $\mu$s  will improve the understanding of the physical phenomenon driving the breathing mode and therefore may help improve thruster design. \cite{Boeuf_2017, Mazouffre_2016, Hara_2019} Of equal significance,  a more accurate characterization can help acquire basic thruster information such as the location of propellant ionization and acceleration, \cite{Vaudolon_2014} the ion speed and the direction of ejection, \cite{Hargus_2001, Mazouffre_2013} the electron transport across the magnetic field of the thruster channel and in the near field, \cite{Meezan_2001, Janes_1966} etc.

The importance of the electron dynamics to the thruster's fundamental operation has stimulated concerted efforts to study it numerically, experimentally, and analytically. \cite{Boeuf_1998, Hara_2014, Lobbia_2010, Mazieres_2022} It has been shown that classical models of electron transport across magnetic lines underestimates the electron current by orders of magnitude.\cite{Dale_2019, Meezan_2001} An anomalous transport mechanism is likely to be accountable for enhancing this current, as results by the measurements, and concurrent experimental and theoretical investigations are required to characterize the relationship between the dominant {\it breathing} oscillations and the anomalous electron transport.

On the experimental side, LIF technique is the preferred means for studying breathing mode dynamics thanks to the rich information that can be obtained with negligible intrusiveness, and good spatial and time resolution. One of the main challenges that remain is the presence of a poor signal-to-noise ratio, combined with an oscillating current of the thruster discharge that lacks strict periodicity. This necessitates a lengthy measurement time without loosing information about the instantaneous phase of oscillation of the thruster current, unless specifically driven to a coherent breathing frequency, as demanded by certain time-resolved LIF (TR-LIF) approaches.\cite{Balika_2013}

Numerous techniques have been developed to implement TR-LIF diagnostics, placing emphasis on different aspects: such as obtaining high resolution in time or ion velocity. Significant results in TR-LIF have been obtained using: photon-counting, \cite{Balika_2013} heterodyne detection, \cite{Diallo_2015} transfer function averaging, \cite{Durot_2016} boxcar, \cite{Young_2016, MacDonald_2012} sample-hold, \cite{MacDonald_2012} and fast switching \cite{Fabris_2015} techniques - a thorough analysis of the various TR-LIF techniques has been documented by C.V.Young and co-workers.\cite{Young_2018}

This work presents a novel technique to study the ion dynamics inside the thruster channel encompassing the near field region. A fully numerical approach for simultaneous measurement of the IVDF relative to a pre-determined set of targeted phase intervals within an oscillation period of the discharge current is proposed and validated by application in a low-power HET. This technique, named phase-resolved LIF (PR-LIF), can significantly reduce the complexity of the diagnostic bench and shorten the measurement time by using parallelized multiple IVDF measurements. The influence of the quasi-periodic behaviour of oscillation of the HET current is counteracted by determining its instantaneous phase of oscillation.

The paper is organized as follows. Section \ref{test facility} briefly introduces the test facility. The LIF set-up arrangement and characteristics are briefly depicted in Section \ref{LIF set-up}. Section \ref{PR-LIF} describes of the PR-LIF technique and Section \ref{results} presents time averaged and PR-LIF results in low power HET. Finally, the conclusions are set out in Section \ref{conclusions}. A more detailed description of the phase extraction method and its limitations is presented in the Appendix (Section \ref{Appendix}).
\vspace{-2em}

\section{\label{test facility}Test facility}
The LIF measurements are conducted in a non-magnetic stainless steel vacuum chamber. Vacuum is obtained by a two-stage cryogenic pump  and a single stage cryogenic panel. The base pressure of the vacuum chamber is as low as $10^{-7}$mbar and increases up to about $10^{-5}$mbar during the thruster operation.

\begin{figure}[hb!]
\centering
\includegraphics[width=0.5\textwidth]{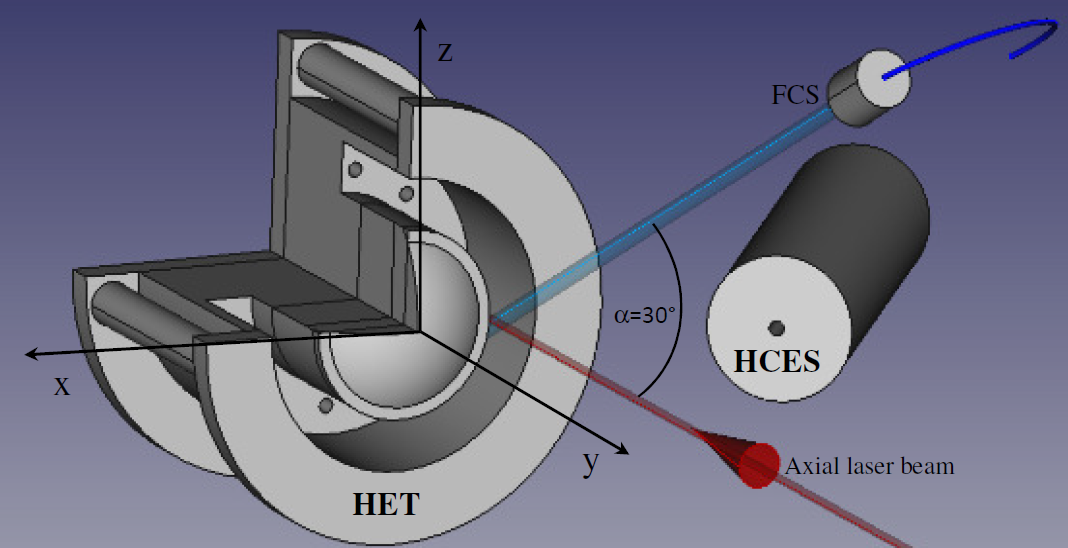}
\caption{Schematics of the HET, the cathode (HCES), and the fluorescence collection system (FCS). The origin of the coordinate system is positioned radially at center of the thruster  ($\hat{x}, \hat{z}$ plane) and longitudinally at the exit plane (along $\hat{y}$).}
\label{HET setup}
\vspace{-2em}
\end{figure}

The thruster is a laboratory model, low-power HET with an outer diameter of 40$\nobreakspace$mm and is mounted on a system with $\hat{x}, \hat{y}$ translation stages (see Fig.\ref{HET setup}) to enable accurate positioning with respect to the measurement point. An overflowing, commercial cathode is used as an electron source. The thruster operating point studied in this work is characterized by an intense breathing oscillation. Details on the thruster electrical parameters are given in 
 Table$\nobreakspace$\ref{tab:OP}. 

\begin{table}
\caption{\label{tab:OP}HET operating parameters.}
\begin{ruledtabular}
    \begin{tabular}{lccr}
    \hline
      \textbf{Parameter} & \textbf{Value} \\
      \hline 
      Anode potential & 200$\nobreakspace$V \\
      Anode mean current & 0.79$\nobreakspace$A \\
      Anode oscillating current & 0.330$\nobreakspace$ A$_{rms}$ \\
      Breathing oscillation frequency & at about 30$\nobreakspace$kHz \\
      \hline
      \vspace{-1em}
        \end{tabular}
 \end{ruledtabular}
 \end{table}

\section{The LIF set-up}
\label{LIF set-up}
The light source used for Xe$\nobreakspace$II excitation at 834.7233$\pm$0.0001$\nobreakspace$nm \cite{Gawron_2008}  (5d$^2[4]_{7/2}\rightarrow$6p$^2[3]_{5/2}$ transition) is a tunable diode laser in a master-and-slave configuration. The selected metastable state is assumed to provide a good representation of the entire ion population.\cite{Konopliv_2023}

\begin{figure}[h]
\centering
\includegraphics[width=0.5\textwidth]{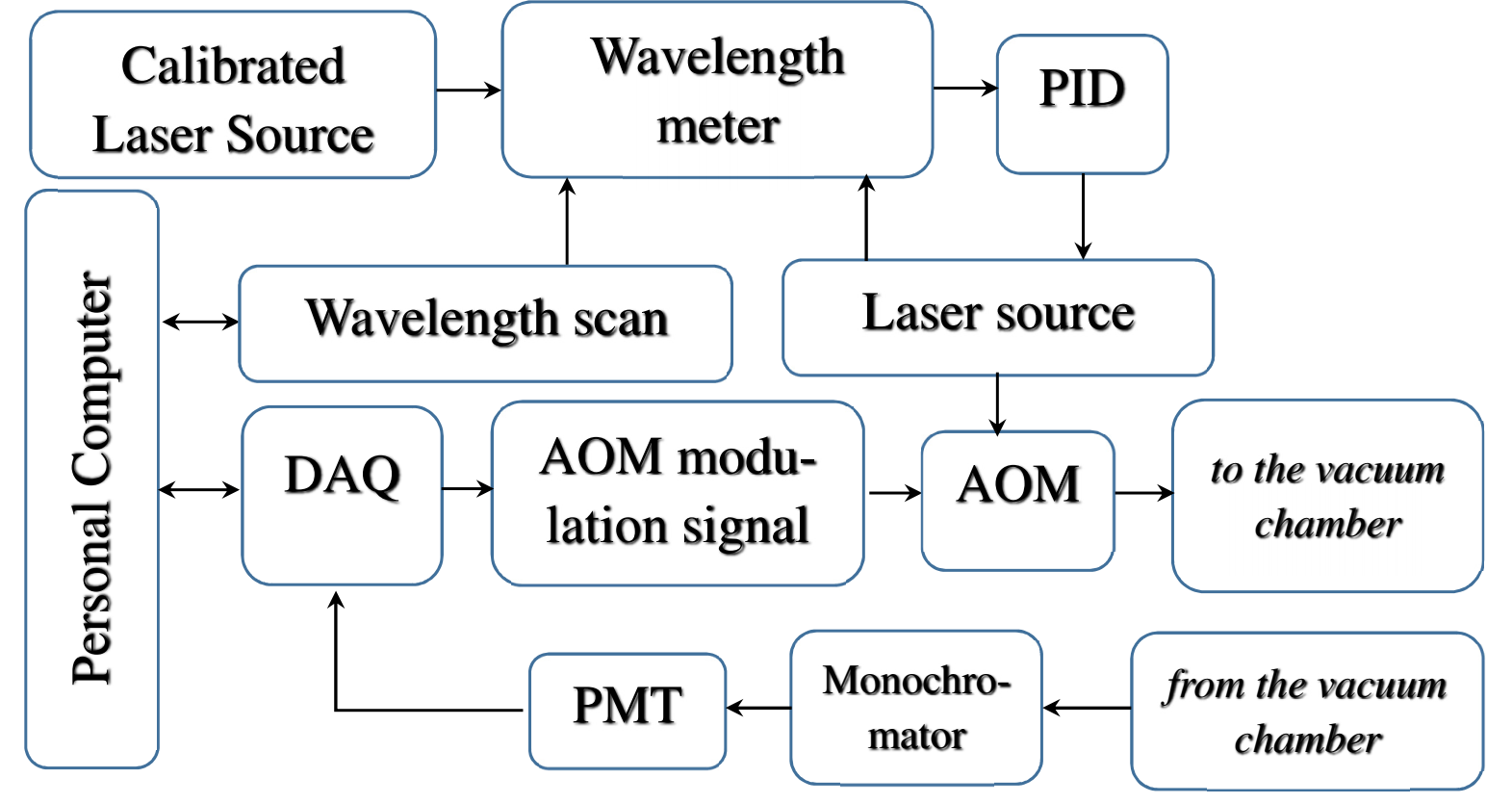}
\caption{Schematic of the LIF set-up: Proportional-Integral-Derivative (PID) controller; acousto-optical modulator (AOM); photo-multiplier tube (PMT); and data acquisition card (DAQ).}
\label{LIFsetup}
\vspace{-1em}
\end{figure}

A schematic of the LIF set-up is given in Fig.\ref{LIFsetup} (a more detailed description can be found in Refs.\onlinecite{Dancheva_2022, Dancheva_2013}). The laser wavelength is locked and scanned using a high accuracy ($\pm\nobreakspace10\nobreakspace$MHz) wavelength meter and a proportional-integral-derivative (PID) controller. A wavelength scan of about 0.05$\nobreakspace$nm typically lasts for about 120$\nobreakspace$sec. The wavelength meter undergoes periodic calibration using a diode laser that is frequency stabilized to the  absorption profile of the Caesium D$_2$ line, ensuring an absolute accuracy of $\pm\nobreakspace2\nobreakspace$MHz. Considering solely the aforementioned errors the uncertainty in determining the ion velocity is not better than $\pm\nobreakspace37\nobreakspace$m/s.

The laser intensity is modulated using the fibre-coupled acousto-optical modulator (AOM) at 200$\nobreakspace$Hz. In general, multiple beams plasma excitation schemes can be realized. \cite{Dancheva_2022}  In this work axial plasma excitation beam is applied (laser power in the tens of mW range) with a beam waist diameter of 0.8$\nobreakspace$mm at the measurement point. The plasma fluorescence signal is collected using an objective coupled to a multi-mode fibre with a view spot diameter of 2.5$\nobreakspace$mm.

The fluorescence signal at 541.915$\nobreakspace$nm (corresponding to the transition 6p$^2[3]_{5/2}\rightarrow$ 6s$^2[2]_{3/2}$) is selected using a grating monochromator. Subsequently, the signal is detected using a photo-multiplier tube (PMT). Time-averaged (TAv-LIF) and PR-LIF measurements inside the thruster channel (central line) and in the near field of the plume are conducted translating the thruster along the $\hat{y}$ direction to produce IVDF maps.

\section{\label{PR-LIF}Description of the PR-LIF method}
 The results presented here encompass several distinct timescales: slower one determined by the laser intensity modulation (on the order of few milliseconds), an intermediate one determined by the period $T$ of the breathing oscillation of the HET current (in the tens of microseconds range), and the fastest one dictated by the microseconds acquisition rate. Some of the TR-LIF techniques, as for example the boxcar one, use fixed time interval $\Delta t$ within the oscillation period $T$ to extract time-dependent LIF profiles, where $\Delta t<T$.
 However, averaging over many oscillation periods $T$ with the consequent increase of the measurement time is usually applied to improve the low signal-to-noise 
 characteristic for the LIF diagnostic. To ensure rigorous time-correlation when examining a quasi-periodic phenomenon across multiple periods, it becomes necessary to apply a stretching or compressing of the data time-series, thereby making it appear nearly periodic. This data manipulation enables a fixed $\Delta t$ to be used to conduct correlation analysis with other system parameters. An alternative approach would be to adjust the $\Delta t$ duration aligned with that of the  current period $T$ of oscillation. The PR-LIF technique, here proposed, is innovative in this respect. More specifically, the PR-LIF technique samples the breathing oscillation at specified phase-intervals ($\Delta\phi$) instead. This approach effectively accounts for the quasi-periodicity and enables its application to naturally oscillating plasma. 
  
 Special care is taken to create a simple and compact set-up, capable of providing maps of the IVDF evolution. The proposed technique also enables simultaneous sampling at different phases of the oscillatory phenomenon under investigation, parallelized PR-LIF.

 For the technique implementation, two signals are simultaneously acquired with a sampling rate of 1$\nobreakspace \mbox{Msps}$: the PMT output (see Fig.\ref{LIFsetup}) and the thruster discharge current. The data flow is arranged in a first-in-first-out-like buffer and the elaboration is performed in data packets. 
 
\begin{figure}[h]
\centering
\includegraphics[width=0.46\textwidth]{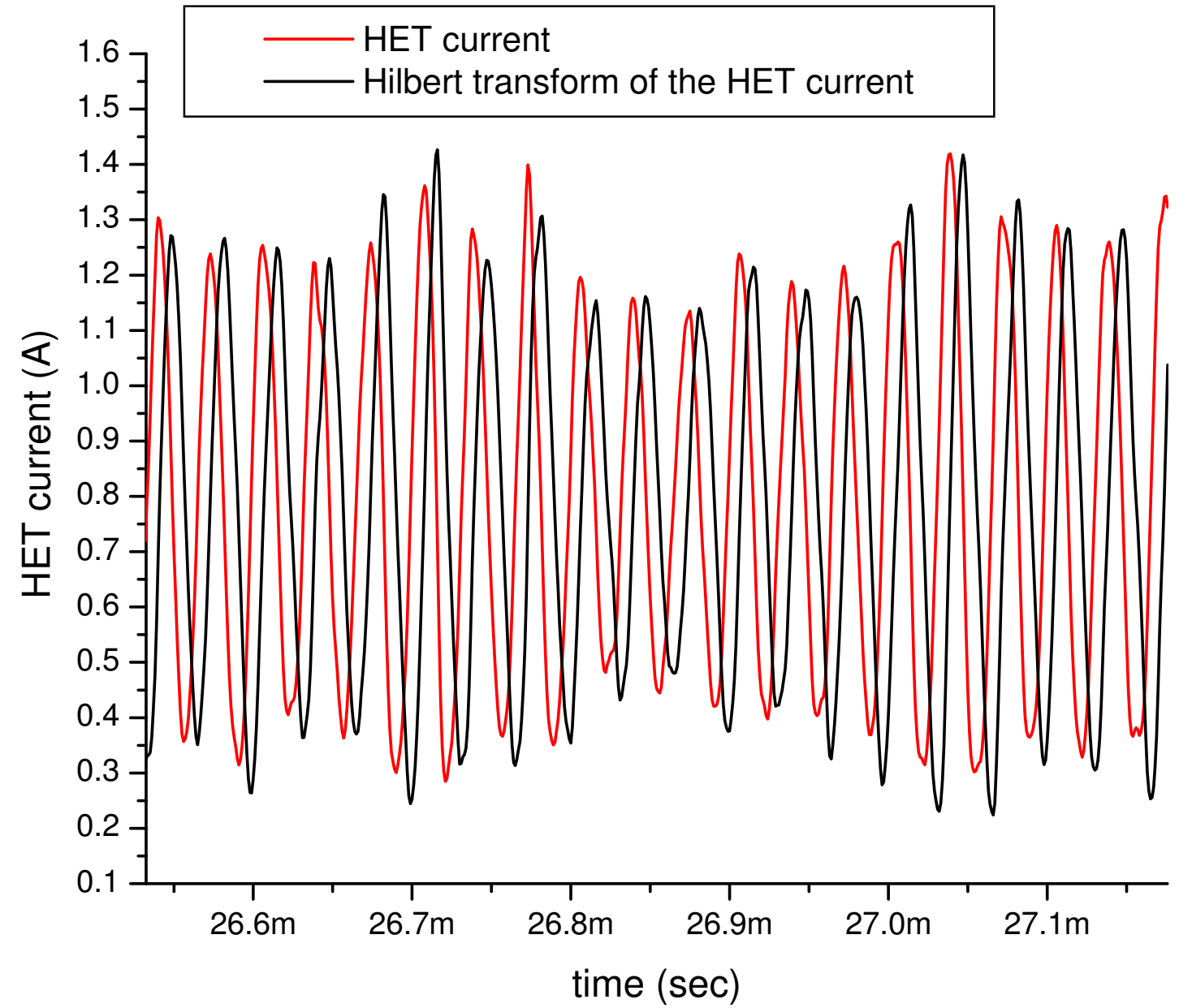}
\caption{Example of the oscillating HET current (black solid line) and its Hilbert transform (red solid line).}
\label{Sin Cos}
\vspace{-1em}
\end{figure}
\begin{figure}[!h]
\centering
\includegraphics[width=0.46\textwidth]{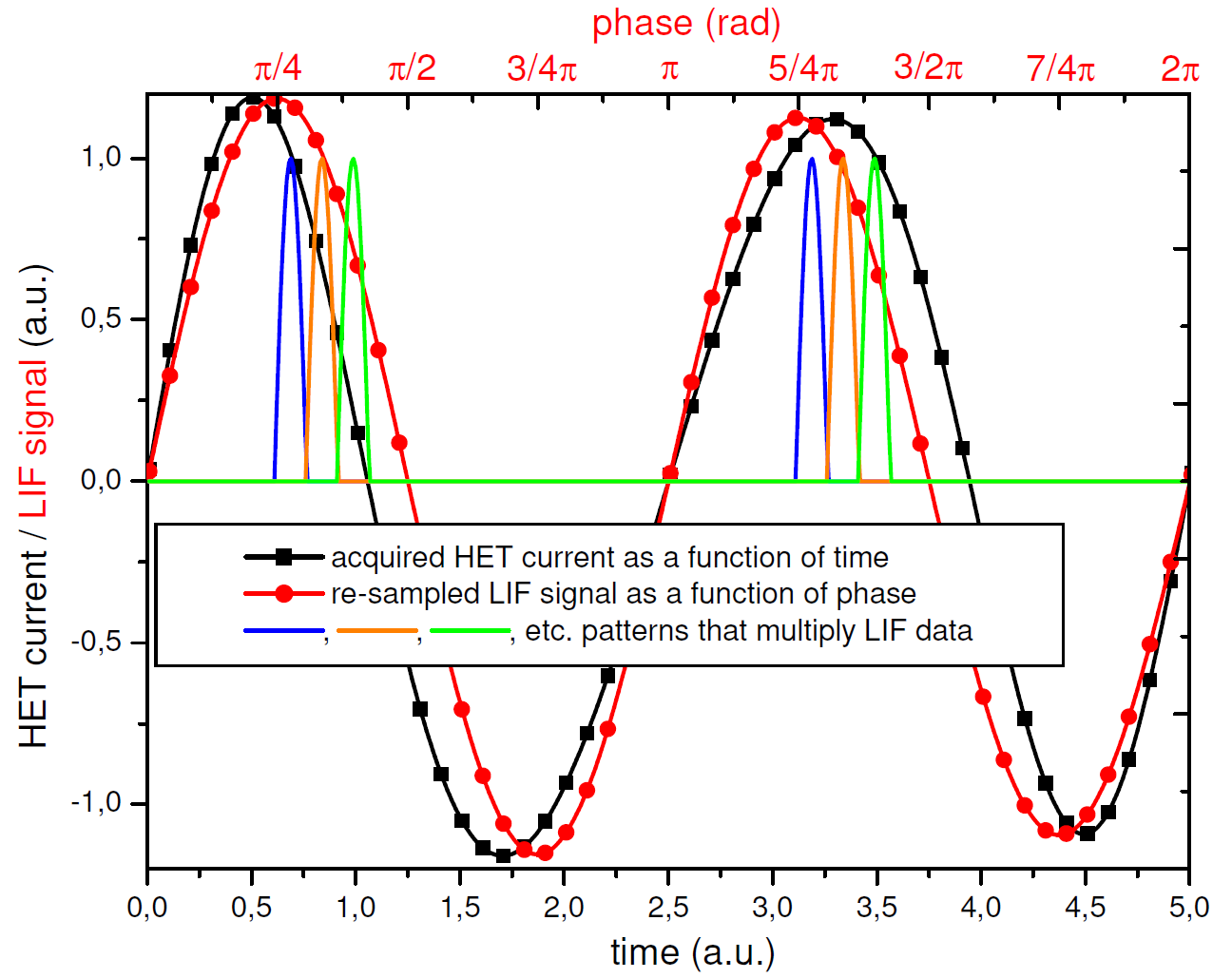}
\caption{Example of the PR-LIF data re-sampling technique, using simulated signals, over two breathing periods. The green, the blue and the orange solid lines represent example of the patterns (Hanning window) used to multiply the LIF data - the shape of the window can go from Hanning to a rectangular one. The number of patterns is determined by the breathing period sampling. The breathing oscillation is now transformed into a periodic one and the following data handling is performed as a function of the phase.}
\label{resample}
\vspace{-1em}
\end{figure}

The first step in the data analysis aims to extract the breathing oscillation instantaneous phase. Hilbert transform is applied to the measured signal $x(t)$ (discharge current) to recover a corresponding imaginary signal $y(t)$, such that the $S(t)=x(t)+iy(t)$ is the analytical signal (see Fig.\ref{Sin Cos}). A more detailed description on rigorous determination of the analytical signal is given in the Appendix. The breathing instantaneous phase $\phi(t)$ is evaluated, according to: $\arg(S)=\atantwo(x, y)$  and the LIF signal is re-sampled (see the example given in Fig.\ref{resample}) at a constant phase rate (typically $\delta\phi=\nicefrac{2\pi}{40}$ having $\Delta\phi=n\delta\phi$, where $n$ is an integer equal to 5 in this work) by means of a linear data interpolation.

\begin{figure}[!h]
\centering
\includegraphics[width=0.47\textwidth]{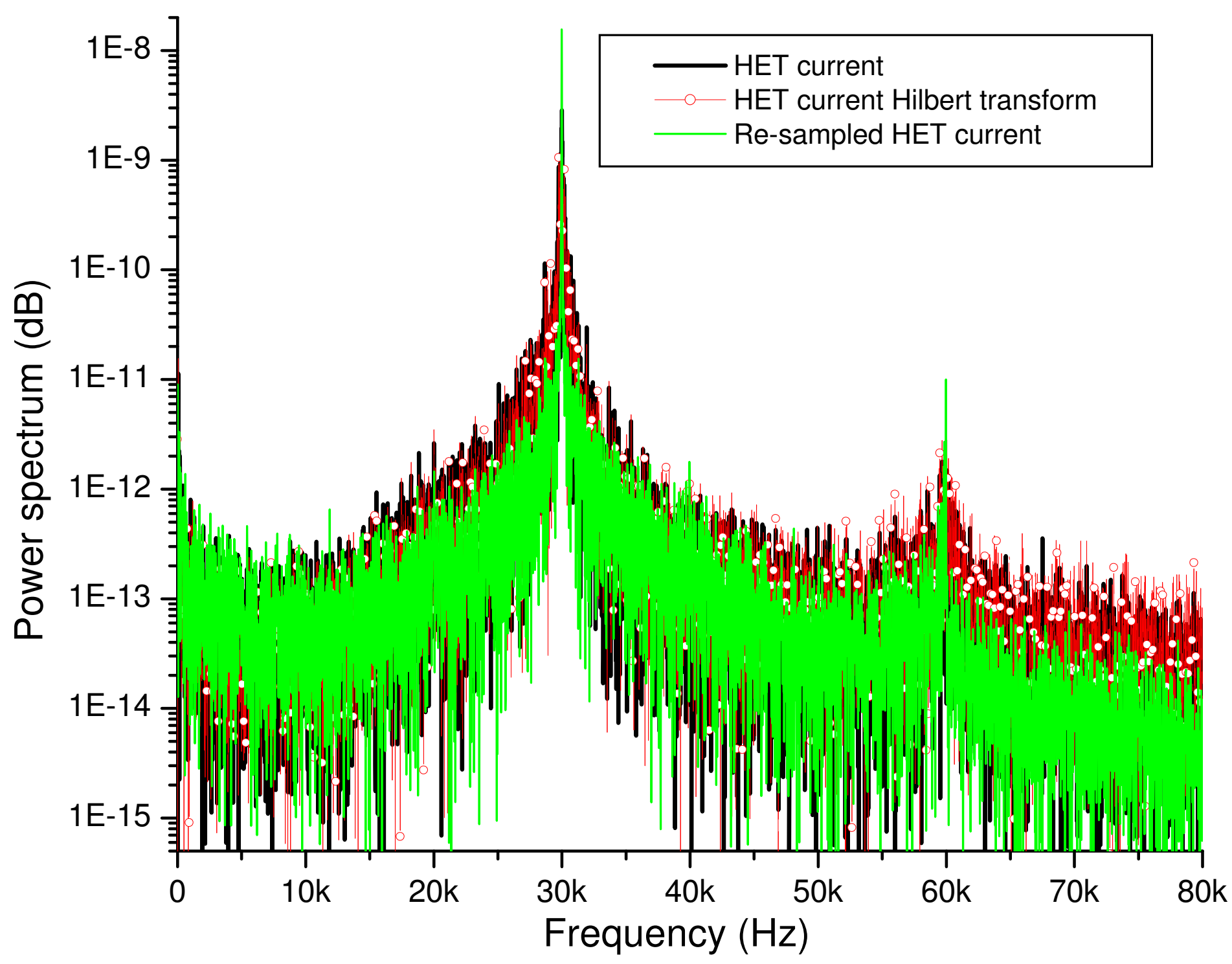}
\caption{Power spectrum of the HET current (black plot), its Hilbert transform (red scattered plot), and the re-sampled one at fixed phase rate $\delta\phi$ (green plot). The power spectrum of the re-sampled HET current is evaluated at sampling time $\delta t=\overline{T}\nicefrac{\delta\phi}{2\pi}$, where $\overline{T}$ is the mean period of oscillation.}
\label{FFT Hilbert}
\vspace{-1em}
\end{figure}
\begin{figure}[!h]
\centering
\includegraphics[width=0.46\textwidth]{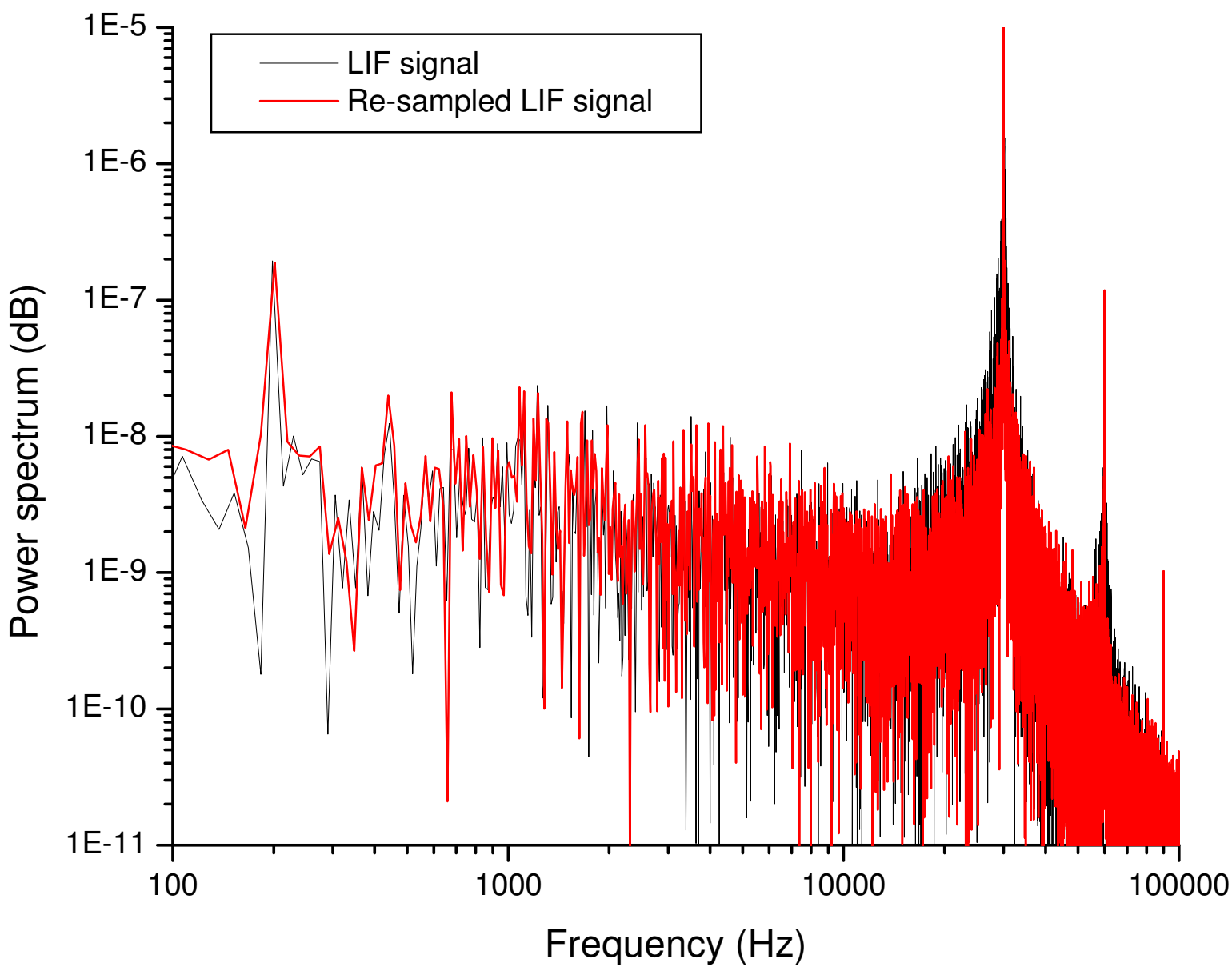}
\caption{The original LIF signal power spectrum (black plot) and the re-sampled one (red plot) at $\delta\phi$. As in Fig.\ref{FFT Hilbert} the power spectrum of the re-sampled LIF signal is evaluated at sampling time $\delta t$.}
\label{FFT LIF}
\vspace{-1em}
\end{figure}

The HET current re-sampling in phase is not strictly required for the PR-LIF analysis. However, this can help verify the transformation of the signal into a periodic form through re-sampling. This is evident from the notable narrowing of its  spectrum, as depicted in Fig.\ref{FFT Hilbert} (green solid line) in comparison to  the spectra of the acquired signal $x(t)$ and its Hilbert transform $y(t)$. It should be noted that the extracted instantaneous phase strictly corresponds to the phase of the fundamental breathing mode. Ongoing research aims to establish its correlation with the characteristic shape of the breathing period, provided such a form exists. It is worth mentioning that the presence of a strong second harmonic can significantly alter the latter's profile.

The breathing period $T$ is now separated in targeted phase intervals $\Delta\phi$ and a set of patterns are prepared (see Fig.\ref{resample}), each operating in a given $\Delta\phi$. The pattern shape can go from rectangular (no windowing), with an amplitude equal to one within the specific $\Delta\phi$, to the Hanning window (Fig.\ref{resample}).

Each LIF data packet (overall time duration of about few tens of msec) is multiplied by the 8 patterns and a numerical demodulation against the laser modulation frequency is simultaneously performed to extract the signal amplitude relative to the targeted phase interval. 

The LIF signal demodulation is performed exclusively by numerical means based on numerical lock-in or simply FFT analysis, with some care to avoid spectral leakage and scalloping losses. \cite{Harris_1978} The AOM modulation signal is numerically generated, synchronously with the data acquisition, which facilitates the implementation of the numerical lock-in. To this aim, an integer number of modulation periods are processed at a time - a single data packet. Special care is taken to keep the time necessary for data manipulation and demodulation within the time necessary for the acquisition of the next data packet, to provide real time output.

Provided that the breathing frequency is much higher than that of the laser intensity modulation, no modification of the LIF signal in the vicinity of the latter (200$\nobreakspace$Hz) is observed  due to the re-sampling in phase (Fig.\ref{FFT LIF}). The analysis presented in this work considers  HET breathing period sampling in 8 targeted phase intervals. The acquisition rate is limited by the time necessary for data manipulation that will allow real time measurements while handling data packets made of 10$^5$ data.

The measurement time at given laser wavelength (numerical demodulation settling time) sets the integration time if no additional numerical filtering is applied - the longer the better is the noise rejection. The laser scan is slow enough to ensure that the wavelength variation, within the settling time, is smaller than the wavelength meter accuracy, thus the ongoing signal amplitude corresponds to the contemporary excitation wavelength. The total measuring time for a single IVDF scan is determined by two factors: the settling time of numeric demodulation and the  scan range of the laser wavelength.

\section{\label{results}TA\MakeLowercase{\bf v}-LIF and PR-LIF Results}
Time averaged LIF measurements are performed to infer the axial IVDF and to validate the PR-LIF technique. 
\begin{figure}[hb!]
\centering
\includegraphics[width=0.46\textwidth]{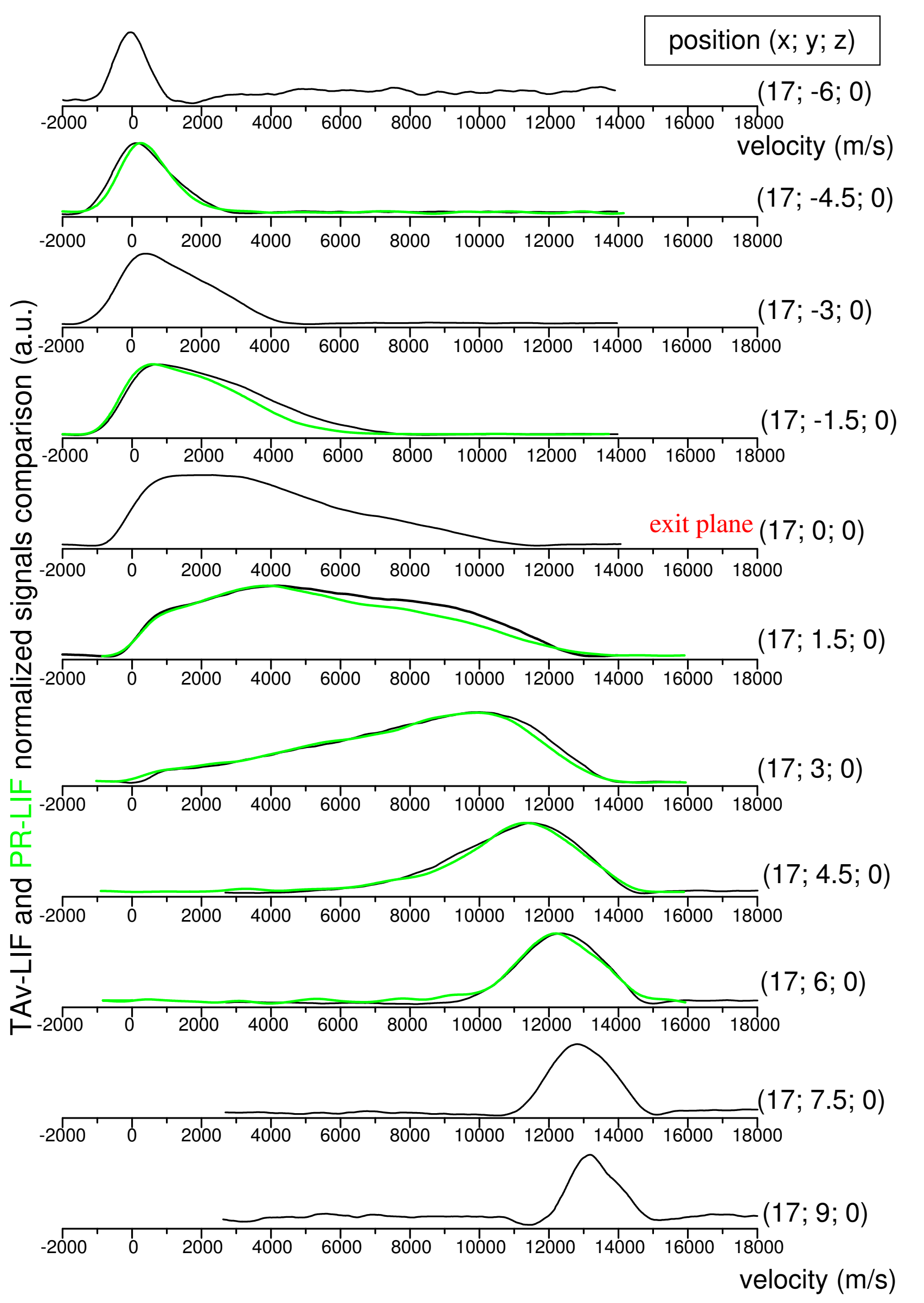}\\
\caption{TAv-LIF traces (black plots) and PR-LIF (green plots) as a function of the axial distance. The thruster exit plane is at y=0.}
\label{Ta to PRLIF}
\vspace{-1em}
\end{figure}

\begin{figure}[hb]
\centering
\includegraphics[width=0.5\textwidth]{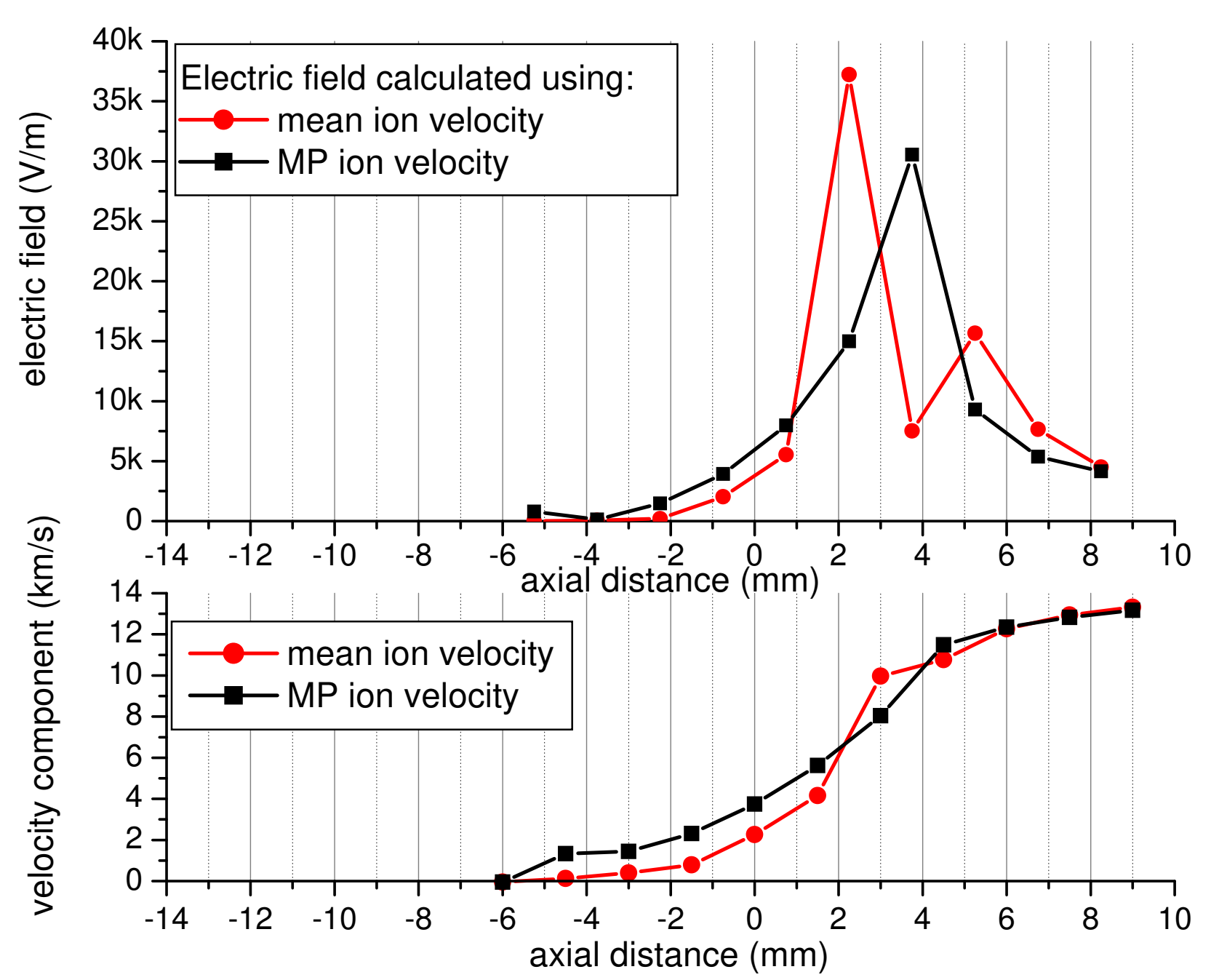}\\
\caption{Axial distribution of the thruster electric field (top), and the most probable (MP) and the mean components of the axial ion velocity (bottom).}
\label{BE}
\end{figure}

Fig.\ref{Ta to PRLIF} shows the spatial evolution of the axial IVDF at the channel central line as a function of the displacement along $\hat{y}$. The most probable velocity and the mean velocity (calculated using the first moment of the IVDF) are derived from the TAv LIF traces (see the bottom panel in Fig.\ref{BE}) and are used to calculate the time-averaged electric field (top panel in Fig.\ref{BE}).

Both the main ion acceleration and the highest velocity dispersion take place few millimetres downstream in a narrow spatial range (see Fig.\ref{BE} and plots from 3 to 7 in Fig.\ref{Ta to PRLIF} starting from the top). The highest LIF intensity (determined by the ion metastable state 5d$^2[4]_{7/2}$ population) observed at about y=-1.5$\nobreakspace$mm is likely to be ascribed to an increased ionization rate in that region.

TAv-LIF and PR-LIF are performed in two distinct test campaigns, where the operating conditions of the thruster are as similar as possible (see Table \ref{tab:OP}). A comparison is shown in Fig.\ref{Ta to PRLIF}. The PR-LIF IVDFs are summed and normalized to the maximum value of the corresponding TAv-LIF trace. A comparison of the TAv and PR LIF traces serves as a valuable validation method. The strong correspondence between the two demonstrates the suitability of the proposed PR-LIF method for characterizing the typical oscillating behavior of a HET.

\begin{figure}[h!]
\centering
\includegraphics[width=0.5\textwidth]{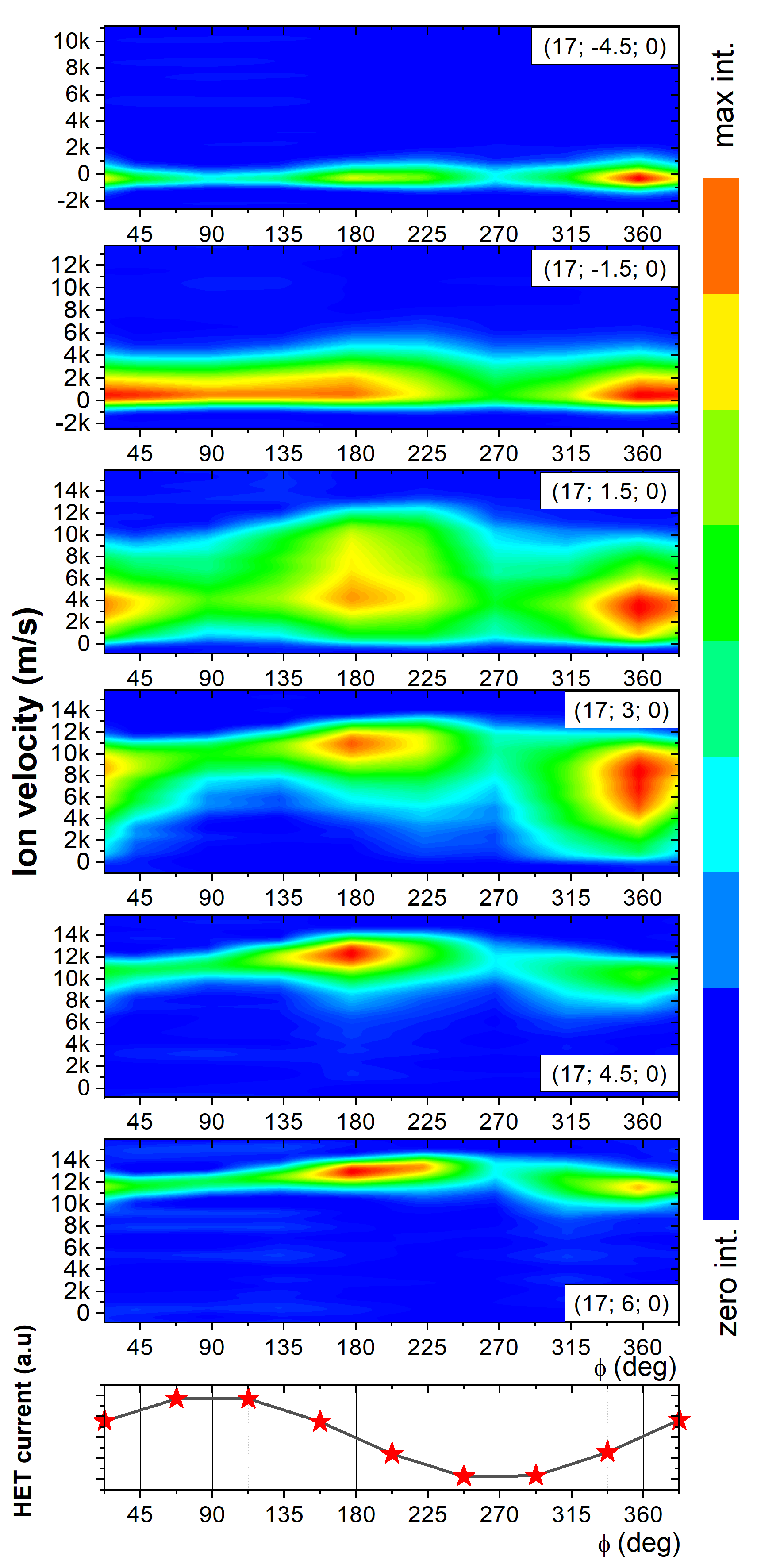}\\
\caption{Maps of the axial IVDF as a function of the HET current oscillation phase $\phi(t)$ and the axial distance (coloured contour plots)  over a single current oscillation period. The (x, y, z) coordinates of the measurement point are given on the right side of each plot. In the lowest plot the center of each targeted phase interval, marked with red star, is given taking into account only the fundamental breathing mode frequency.}
\label{PRLIF}
\end{figure}

The phase lag introduced by both the LIF and HET current registration systems is characterized as a function of the frequency. The PMT signal is converted into voltage using a buffered transimpedance low-pass amplifier with a cut-off frequency of 230$\nobreakspace$kHz and the thruster current is monitored using a current probe (bandwidth of 100$\nobreakspace$kHz). Both signals are acquired using an anti-aliasing  filter (with a cut-off frequency of 110$\nobreakspace$kHz). The presented PR-LIF measurements are adjusted by accounting for the relative phase lag at the breathing frequency.

\begin{figure}[h]
\centering
\includegraphics[width=0.5\textwidth]{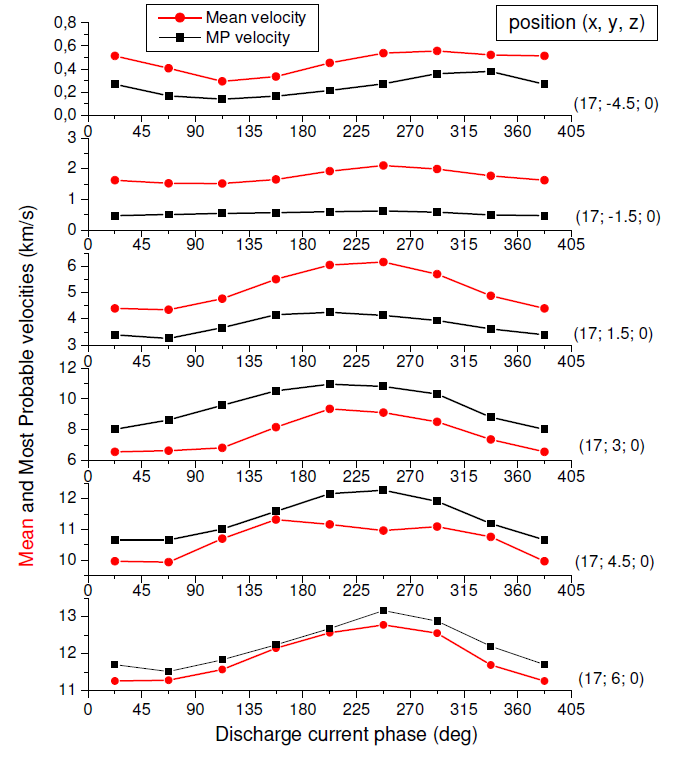}\\
\caption{The dynamic of the axial IVDF mean (red circles) and most probable (black squares) velocities as a function of the HET oscillation phase and the axial distance.}
\label{velo PRLIF}
\end{figure}

PR-LIF measurements are taken at assigned axial positions. The contour maps presented in Fig.\ref{PRLIF} show LIF signals, normalized to the maximum value, as a function of the axial distance and thruster current phase of oscillation $\phi$(t). The maps of the most probable and the mean velocity components are plotted in Fig.\ref{velo PRLIF}. A slight modification of the velocity distribution can be observed at axial distances where no significant acceleration is present (see for example the trace at y=-4.5$\nobreakspace$mm).  The highest ion dynamics is observed (in correspondence with the peak of the electric field)  about $2\div3\nobreakspace$mm downstream from the exit plane. 

\begin{figure}[h]
\centering
\includegraphics[width=0.5\textwidth]{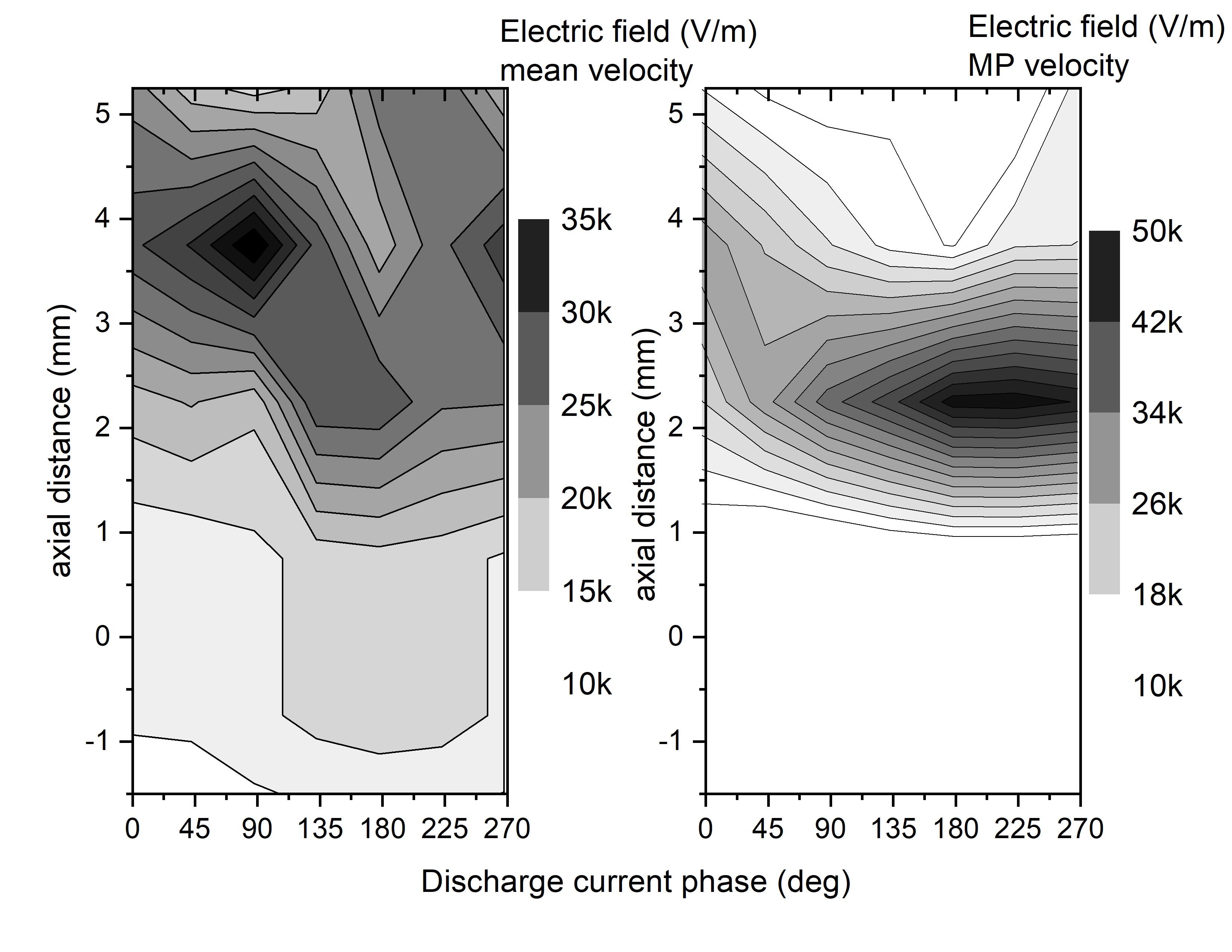}\\
\caption{Electric field dynamics over a single oscillation period evaluated using both the mean (left) and the MP (right) ion velocity. Neither data smoothing nor interpolation is applied prior to electric field calculation.}
\label{EF PRLIF}
\end{figure}

Using both most probable and mean ion velocities, the time dependent axial electric field can be roughly estimated (see Fig.\ref{EF PRLIF}) with the assumption of a steady-state, 1D, collisionless plasma in $\hat{y}$ direction with uniform ion density. \cite{Young_2018}

It is inferred that intense electric field oscillation takes place during the discharge current downward ramp with higher amplitude variation in case of the most probable velocity. A thorough description of the breathing phenomenon and the resulting electric field dynamic can be found elsewhere and it will be briefly summarized here.\cite{Fife_1997, Barral_2009, Boeuf_1998, Young_2018}

Higher HET currents implies higher ionization rates occurring at an increased density of the neutrals in the region of strong magnetic field. However, the higher the neutral density, the lower is the plasma resistivity, with the consequence of a smoother and further from the anode electric field. Analogous considerations can be done for the opposite case (minimum current), where less neutrals are present (no avalanche), hence a higher plasma resistivity is expected resulting in a closer to the anode electric field. The behaviour of the electric field, recovered from the most probable velocity in Fig.\ref{EF PRLIF}, is in good agreement with the physical picture described above.

\section{\label{conclusions}Conclusions}
A fully numerical method is applied to investigate the dynamics of the IVDF in a HET, operating under the conditions of pronounced breathing oscillations. Conducting phase-resolved measurements an average over many breathing oscillation periods is performed, despite the quasi-periodic nature of the oscillation. Provided that the  phase of the thruster current is rigorously recovered, the proposed technique enables the determination of ion velocity with low uncertainty, preserving a good time/phase resolution, applying lengthy measurements. The method can be applied to either driven or spontaneous HET current oscillations. 

The numerical data management significantly reduces the complexity of the diagnostics set-up, as no additional instrumentation is required for its implementation. Real-time IVDF measurements are conducted by sampling the breathing period in a limited number of points. An important advantage of this technique is the significant reduction in measurement time made possible by parallelized analysis across a specific number of points within the oscillating period. Increasing the sampling rate improves the accuracy of the breathing period sampling, although it can pose challenge for real-time output. Different sets of re-sampling/averaging parameters can be selected either to reduce the computational time and aim for real-time output, or to improve the phase resolution and the signal-to-noise for better performance. Accomplishing an offline analysis would allow for a more accurate sampling of the phenomenon's period and provide the opportunity to compare with other TR-LIF techniques, such as the transfer function averaging method.

The PR-LIF is applied here in a low power Hall effect thruster in the case of axial IVDF mapping. A comparison with the TAv-LIF shows very good coincidence with the corresponding PR-IVDFs. Both PR-IVDFs and electric field trends are monitored so as to provide a valuable tool to better understand, model, and characterize an HET thruster in oscillatory regime of operation.

\section{Appendix}
\label{Appendix}
\setcounter{equation}{0}
\renewcommand{\theequation}{A\arabic{equation}}
A real, measurable, quasi-periodic signal, such as the thruster current, in the regime of significant breathing oscillation, can be described as :
\begin{equation}
x(t)=A(t)\cos(\phi(t)),
\end{equation}
where A(t) and $\phi$(t) are respectively, the amplitude and the phase. The pair $\{A(t), \phi(t)\}$ can be unambiguously determined defining an analytical signal:
\begin{equation}
S(t)=x(t)+iy(t)=A(t)e^{i\phi(t)}
\end{equation}
by adding an imaginary part $y(t)$ to the measured one $x(t)$. The amplitude and the phase can be found according to: \cite{Vakman_1996}
\begin{equation}
\begin{split}
    A(t)=\sqrt{x^2+y^2}\\
\phi(t)=\arg(S).
  \end{split}  
  \label{fi}
\end{equation}
The imaginary part of the signal can be derived by the real one using the Hilbert transform $\mathcal{H}$[x(t)]:
\begin{equation}
\mathcal{H}(x(t))=\frac{1}{\pi}\int_{-\infty}^{+\infty}\frac{x(\tau)}{t-\tau}d\tau
\end{equation}
that should satisfy the following conditions: \cite{Vakman_1996, Chavez_2006}
\begin{itemize}
\item Amplitude continuity and differentiability: small perturbation to the signal $x(t)$ induce small change of $A(t)$:
\begin{equation}
    \mathcal{H}[x+\delta x]\rightarrow\mathcal{H}[x],\nobreakspace \text{if} \nobreakspace \lVert \delta x\rVert\rightarrow 0.
    \label{first}
\end{equation}
\item Phase dependence of scaling and homogeneity: scaling the signal by a positive constant $c$ should have no effect on the instantaneous phase and its derivative:
\begin{equation}
\mathcal{H}[cx]=c\mathcal{H}[x].
\label{second}
\end{equation}
\item Harmonic correspondence: for any constant $a>0$, $\omega$>0:
\begin{equation}
    \mathcal{H}[a \cos(\omega t+ \Phi)]=a \sin (\omega t + \Phi).
    \label{third}
\end{equation}
\end{itemize}
It has been shown that the condition $A(t)\cos(\phi(t))+iA(t)\mathcal{H}[\cos \phi (t)]=A(t)\exp(i \phi (t))$ 
is verified for narrow-band signals. \cite{Delprat_1992} A signal is considered narrow-band if the relative variation of its amplitude $A(t)$ is slow when compared with the phase variation, namely the ratio: 
\begin{equation}
    b=\frac{\left|\frac{d\phi(t)}{dt}\right|}{\left|\frac{1}{A(t)}\frac{dA(t)}{dt}\right|},
\end{equation}
has high mean value $\left<b\right>$, and its Hilbert transform satisfies the conditions given by Eqs.\ref{first}, \ref{second}, and \ref{third}. In addition, a single center of rotation is expected for narrow-band signal trajectories on the complex plane. 

\begin{figure}[h]
\centering
\includegraphics[width=0.5\textwidth]{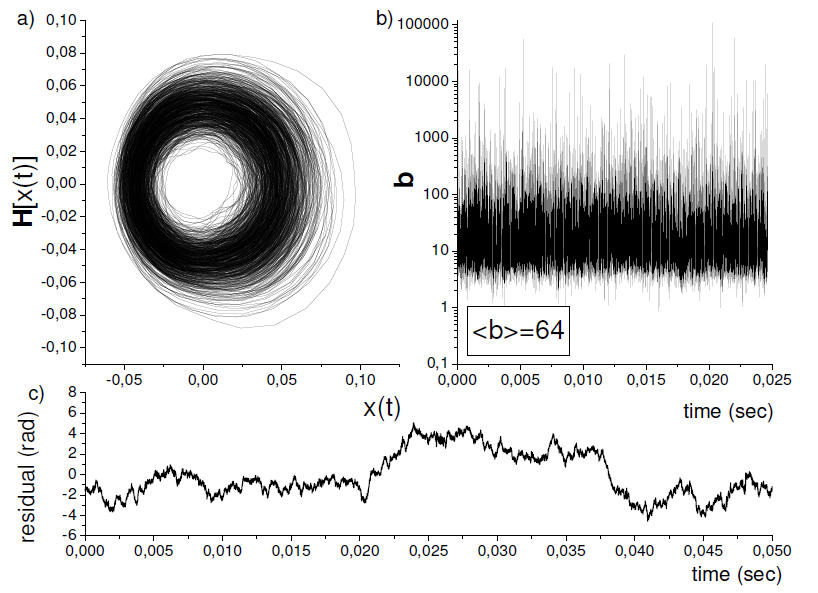}
\caption{Trajectories of the measured HET current $x(t)$ as a function of its Hilbert transform $y(t)$ on the complex plane a) and the ratio $b$ as a function of time  b). The phase deviation from a linear one as a function of time is given c).}
\label{Hilbert}
\end{figure}
The analysis of the typical HET oscillation current shows that it can be determined as a narrow-band signal presenting a single center of rotation and  having the mean value $\left<b\right>$ typically of the order of 60 (see Fig.\ref{Hilbert}). 

The Eq.\ref{fi} is used to calculate the instantaneous phase of oscillation of the thruster current $\phi(t)$ and Fig.\ref{Hilbert}c shows its typical variation from a linear one due to the quasi-periodic nature of the oscillation phenomenon. The typical standard deviation of the instantaneous frequency $\nicefrac{\dot\phi (t)}{2\pi}$, is of about 4.7$\nobreakspace$kHz.

\section{Acknowledgements}{This work has received funding from the European Union’s Horizon 2020 research and innovation programme under grant agreement No 101004331. The authors would like to thank Patricia Nugent for revising the English of the manuscript.}

\section{Data availability statement}
{The data that support the findings of this study are available from the corresponding author upon reasonable request.}

\bibliography{PR_LIF_RSI}
\bibliographystyle{ieeetr}

\end{document}